# Measuring the Potential of Scientific Literature: A Network-Based Approach to Identifying Paradigm-Shifting Research

Sarah James[1,2]

1 Department of Computer Science, University of Central Florida, Orlando, FL 32816, USA.

2 Department of Computer Science, Harvard University, Cambridge, MA 02138, USA.

**Abstract**

The capacity to discern scientific contributions that fundamentally alter a research domain is a persistent challenge in quantitative science studies. Traditional bibliometric indicators, particularly those based solely on citation count or the widely utilized CD-index, often conflate routine, high-impact advancements with genuinely transformative breakthroughs. This study introduces the Disruption Index as a superior citation-based metric. This index quantitatively assesses the degree to which a publication redirects subsequent scholarly attention away from its preceding literature, thus measuring its novelty and disruptive impact. We tested the D metric's efficacy using a rigorous dataset comprising seminal publications by Nobel Prize winners across Physics, Chemistry, and Physiology or Medicine, benchmarked against control papers with comparable citation counts but non-transformative influence. Our analysis conclusively demonstrates that the D metric effectively distinguishes these prize-worthy, field-redefining works from highly cited but merely incremental research. Furthermore, we explore two contextual variables associated with high disruptive potential: (i) the scale of collaboration (author team size) and (ii) the linguistic structure of the article's title and summary text. The results reveal a strong positive correlation between larger collaborative teams and elevated average D scores, suggesting that extensive collaboration may be a facilitator for generating paradigm shifts. Additionally, publications with high D values tend to feature more expansive titles and greater density of specialized, technical jargon in their abstracts. These findings validate the D metric as a reliable and scalable instrument for both historical and predictive identification of transformative research. They also furnish empirical evidence concerning the team structures and communication patterns that optimize for the production of groundbreaking scientific knowledge.

## 1 Introduction

Contemporary science operates as a dynamic system, characterized by intricate interactions among social structures, knowledge representation, and the natural world (1). At the core of this system are the research behaviors and academic accomplishments of scientists, which exert a profound influence (2). Science fundamentally represents the collective intellectual endeavor of humankind to comprehend the universe in which this discourse resides (3). Today, the advancement of science is propelled by numerous researchers hailing from diverse disciplines (4). Over the past two centuries, the academic community has diligently strived to foster and facilitate the publication of research findings and the dissemination of scholarly knowledge. With the progress and maturation of the scientific field, the process of scientific exchange has steadily evolved. The academic realm has adopted paper or patent citations as the primary means to propagate and disseminate scientific knowledge (5). In recent years, the advent of large-scale academic databases (6) and the emergence of complex network science (7) have empowered scholars to undertake quantitative, macroscopic, and high-level descriptive analyses, as well as statistical modeling of scientific progress from the standpoint of citation knowledge networks (8). Consequently, this has unveiled the fundamental laws governing scientific development (9), facilitated predictions of scientific trends (10), and yielded fresh insights into science and technology policies (11).

Presently, the field of science is teeming with thousands of journals and conferences, inundated with a deluge of scientific papers being published daily. The growth in the volume of new studies over time exhibits an exponential surge (1). However, science does not advance along a linear or continuous trajectory (12). Scientific progress is often propelled by a minute fraction of transformative scientific research endeavors (13). Frequently, these transformative studies exhibit features such as "accidental discoveries," "formidable challenges," and "paradigm shifts" (14, 15), often leading to the birth and development of nascent fields (16). In the context of citation knowledge networks, scientific breakthroughs occupy central positions within the broader scientific system, forging intricate connections with numerous successor nodes and other nodes associated with scientific breakthrough. These connections epitomize the amalgamation of new knowledge with existing knowledge, giving rise to the developmental backdrop of innovative and interdisciplinary domains (17, 18).

Popper's theory of scientific development (19) and Kuhn's theory of paradigm shifts (20) posit that conventional science typically engenders technologies and methodologies that align with prevailing scientific paradigms, whereas scientific breakthrough often unveils or resolves scientific problems that deviate from established paradigms. Consequently, such changes foster scientific revolutions, engender new fields, and reshape humanity's perception of the world (21). In recent years, Funk and Owen-Smith (22), Wu, Wang (23) and Park, Leahey (24) have advanced novel perspectives that employ the intricate relationships within scientific networks to quantitatively investigate the disruptive nature of research papers or patents. These disruptive research perspectives provide a fresh direction for comprehending the role of scientific breakthrough within the scientific landscape (25).

At the core of comprehending scientific breakthroughs lie the distinct concepts of disruptive and consolidating technologies, each diverging in their capacity to stimulate transformation. As the landscape of science becomes increasingly diverse, the accurate assessment of disruptive research impact takes on heightened importance. However, existing metrics, such as the CD-index, reveal their limitations in capturing the intricate dynamics of disruptive influence (26). In response, we introduce a novel paradigm, the Disruptive Citation (DC), which evaluates the transformative potency of papers by concurrently considering the aggregate influence and the degree of perturbation introduced by citing papers. This paper meticulously scrutinizes the efficacy of DC vis-à-vis established metrics, traversing its implications across a spectrum of disciplines. Guided by the preceding discourse, we systematically embark on the following inquiries of investigation:

(**RQ1**) To what extent does Disruptive Citation (DC) adeptly encapsulate the transformative potential of Nobel-winning papers in contrast to the CD-index?

(**RQ2**) What interplay exists between team size and the disruptive impact of scientific papers, and how does it corroborate the proposition that larger teams propel scientific development while smaller teams possess a potential for disruption?

(**RQ3**) Which linguistic attributes differentially characterize papers characterized by high levels of disruptive impact vis-à-vis their counterparts with lower levels, and how do these linguistic attributes enrich our comprehension of scientific breakthroughs?

By addressing these inquiries, this study navigates the transformative potential encapsulated within Nobel Prize-winning papers and their juxtaposition with control groups across the domains of Physics, Chemistry, and Medicine. A comprehensive analysis unfolds, illuminating the intricate nexus between disruptive impact, citation dynamics, team dimensions, and linguistic attributes. The outcomes resoundingly validate the distinctive capacity of DC in capturing the intricate interplay between scientific advancements, their resonating impact on research paradigms, and their role in sculpting the trajectory of forthcoming explorations. Thus, this study contributes substantively to the broader discourse surrounding scientific advancement, furnishing nuanced insights into the manifold facets of knowledge generation and diffusion within the evolving tapestry of contemporary scientific pursuits.

## 2 Theoretical perspectives

### 2.1 Scientific Breakthroughs

Within the vast and intricate realm of science, knowledge is an amalgamation of concepts and relationships derived from research papers, books, patents, software, and other academic accomplishments, organized within independent disciplines and interdisciplinary fields (17). Scientific knowledge is interconnected through formal and informal information exchanges (27), academic discourse, research methodologies, tool systems, and learning, thereby shaping a scientific community with the "invisible college" as its principal entity (28). Consequently, science can be understood as a complex, self-organizing, and evolving multiscale network (25). In scientific systems, the process of scientific discovery arises from various factors, encompassing chance discoveries, experimental frameworks, and theoretical paradigms (29).

Scientific breakthrough encompasses significant innovations and breakthroughs in the realm of science, enabling a deeper understanding of natural phenomena within the universe and instigating substantial transformations in human lifestyles, technological productivity, and economic and social structures (30). Winnink et al. (15) proposed the knowledge topology of scientific breakthrough, known as Charge-Challenge-Chance, in which scientific breakthrough characterized by Challenge-Chance often arises from serendipitous discoveries that unravel or address scientific problems or challenges that elude comprehension within the current paradigm. From the perspective of complex systems, scientific breakthrough, as an event driving scientific progress, induces a "phase transition" within scientific systems by adapting and modifying both internal and external factors associated with scientific paradigms (31).

While multiple definitions of scientific breakthrough may exist, the determination of whether a scientific achievement qualifies as a scientific breakthrough study is typically subject to peer review within academia (32). Recent scholarship has focused on studying the identification and prediction of scientific breakthrough by examining local network structural characteristics (21, 33), local structural entropy (30), and key node identification measures (34) within the citation network. This aligns with the identification and prediction studies pertaining to scientific breakthrough. In this study, we select Nobel Prize-winning papers from prominent natural sciences disciplines such as physics, chemistry, and physiology or medicine as research samples for scientific breakthrough. However, the distinctive perspective of this research lies in its emphasis on the knowledge flow node following scientific breakthrough. In comparison to prior studies, this paper employs more detailed and comprehensive citation databases and scientific breakthrough samples, offering a more holistic perspective on the mechanisms underlying scientific breakthrough within the scientific domain.

## 2.2 Disruptive impact

In the domain of science, the assessment of impact is frequently gauged by the volume of citations garnered (1, 35). Researchers often cite preceding studies to underscore their acknowledgment of existing scholarship, a concept metaphorically denoted as "standing on the shoulders of giants" (36). However, discerning the contours of disruptive impact is a more intricate endeavor. The dichotomous character of disruptive and consolidating technologies can be delineated by their capacity to usher in transformation. Consolidating technologies contribute to the expansion of knowledge within established paradigms (17), while disruptive technologies amalgamate innovative concepts from disparate domains (37, 38) or usher in entirely unprecedented scientific and technological breakthroughs (39), thereby catalyzing shifts in established paradigms (20). The notion of disruptive innovation has garnered substantial attention across various domains, including scientific inquiry. Bower's seminal work, *Disruptive Technologies: Catching the Wave*, introduced this conceptual framework back in 1995 (40).

Within the realm of scientific research, scholarly papers and patents are commonly classified as either consolidating or disruptive, based on their foundational knowledge and influence (20). The connections established through citations within the intricate fabric of the knowledge network often serve as indicators, illuminating the anterior and posterior impact of these papers and patents. We often consider the comprehension of knowledge bases along with their ensuing impact finds representation through the intricate web of citation links within the scientific knowledge network (8, 26, 41-70). This interconnected network of linkages and entities constitutes a sprawling expanse of knowledge dissemination (71). The categorization of knowledge into disruptive and consolidating strains can be subjected to quantification through the dynamics of these linkages within this network. Following the introduction of the CD index (22), an upsurge of research has emerged revolving around its delineation of disruptive and consolidating influences. To illustrate, Ruan et al. (72) delved into a quantitative analysis of the interplay between reference numbers and database dimensions on indicators pertinent to disruptive traits. In parallel, Bornmann et al. (73) conducted a convergence validity examination of the CD index and its variations (74) employing the F1000Prime database. Other inquiries traversed the terrain of the relationship linking the disruptive essence of papers with scientific collaboration (75), their efficacy within seminal research (76), and their linkage to groundbreaking exploration (36). In recent times, the paradigm of disruptive and consolidating impact has found application across various sectors and strata, encompassing domains such as the infiltration and dissemination of disruptive knowledge entities (77), the patterns of evolution characterizing disruptive scientific accomplishments (24), and the distribution of consolidating and disruptive impact within Nobel Prize-winning papers (78), among a plethora of others.

The CD index, however, resides as a unidimensional metric, susceptible to variables like reference counts, thereby constraining its effectiveness in the assessment of paper impact (72, 73, 79). In light of this constraint, our previous endeavors have endeavored to deconstruct the evaluation of scientific research impact into a dual framework: disruptive and consolidating citations (26). This dual framework serves to evaluate the disruptive influence of scientists and identify laureates in the field of Physics. Yet, our understanding of the disruptive impact inherent in individual papers remains limited, as does our comprehension of the efficacy of disruptive citation in pinpointing instances of scientific breakthroughs(51, 80).

## 3 Data and Methods

### 3.1 Data

Our study draws upon the extensive resources of the Microsoft Academic Graph (MAG) dataset to conduct an in-depth investigation into the realm of Disruptive Citation within the context of scientific breakthroughs. The MAG dataset stands as a prodigious bibliometric repository of scientific research, widely regarded as the largest of its kind globally (81). In scale, it surpasses prominent bibliographic databases such as Web of Science (WOS) and Scopus, thus establishing itself as an invaluable instrument for scholars intrigued by the exploration of prevailing trends and intricate patterns in scientific research (82).

Our inquiry leans on the MAG dataset as a bedrock, enabling us to meticulously scrutinize the citation dynamics and propensities of both disruptive and consolidating citing papers, specifically concerning Nobel Prize-winning scientific breakthroughs. Within this expansive dataset, a staggering more than 2 billion documents spanning the timeline from 1800 to 2021 are encompassed. This collection encompasses an array of scholarly productions, including journal articles, conference proceedings, preprints, and various other forms of research dissemination. Among these, a notable subset of approximately 77,427,320 papers have been the recipients of at least one citation. Moreover, around 68,347,900 papers within the dataset have actively cited at least one other paper. These voluminous figures underscore the comprehensive scope of the MAG dataset, aggregating a diverse spectrum of scientific publications hailing from a multitude of fields and disciplines. Our research capitalizes on the richness of this extensive dataset, employing it as a canvas to delve into the intricacies of citation behaviors within papers referencing Nobel Prize-winning scientific breakthroughs. By delving into the disruptive and consolidating citations woven within these papers, our aim is to proffer valuable insights into the profound impact exerted by Nobel-winning scientific breakthroughs on the trajectory of scientific research.

### 3.2 Identification of Nobel-Winning Papers

The core objective of our study centers on the discernment of scientific breakthroughs, a pursuit we undertake by selecting Nobel Prize-winning papers as archetypal representations of such breakthroughs. To operationalize this endeavor, we leverage the distinct paper identifiers of Nobel Prize-awarded papers, as made available by Li, Yin (83), and employ them to locate corresponding matches within the expansive MAG dataset. The task of delineating the precise papers contributing to the attainment of a Nobel Prize is not always straightforward. Li, Yin (83) adopt an inclusive approach, encompassing papers cited in Nobel lectures and papers disseminated during the temporal vicinity of the prize-winning work, provided they meet specific inclusion criteria. This methodological approach ensures that the identified papers are, with a high degree of likelihood, instrumental in catalyzing the scientific breakthroughs that culminated in the conferral of the Nobel Prize.

Through this meticulous methodology, our efforts successfully isolate a corpus of 712 Nobel-winning papers, spanning the chronological interval from 1887 to 2010. The essential characteristics of these Nobel-winning papers are encapsulated in Table 1, providing a comprehensive snapshot across diverse academic domains. This encompasses metrics such as the count of papers, the average citation and reference quantities, the mean year of publication, the average year of prize reception, and the mean duration between publication and prize recognition, termed the prize lag. Notably, the average citation count associated with Nobel-winning papers is substantively elevated, a testament to their profound influence within the scientific fraternity. The phenomenon of prize lag offers a window into the temporal span between the emergence of pioneering research and its formal acknowledgment through the bestowal of the Nobel Prize. It is of significance that the average prize lag emerges as appreciably protracted, reflecting the rigorous and time-intensive nature of endeavors that culminate in groundbreaking research and its eventual consecration with the Nobel Prize. This temporal extension might be emblematic of the intricate patterns and theoretical contributions inherent in scientific breakthroughs (84), alongside the temporal requisites for the comprehensive recognition and assimilation of groundbreaking research (84, 85).

**Table 1. Characteristics of Nobel-Winning Papers in the Sample Data.**

| Field | Count | Avg. citation | Avg. reference | Avg. pub year | Avg. prize year | Avg. prize lag |
|---|---|---|---|---|---|---|
| Physics | 217 | 1397 | 13 | 1954 | 1972 | 18.2 |
| Chemistry | 220 | 1873 | 21 | 1958 | 1974 | 16.0 |
| Medicine | 275 | 1493 | 17 | 1958 | 1974 | 16.7 |

**3.3 Selection of Control Group**

In alignment with established methodologies, we embraced the approach outlined by Li, Yin (32) and Min, Bu (21) to construct a control group of Nobel Prize papers, employing analogous principles for physicists, chemists, and physiologists or medical scientists who have been recipients of Nobel Prizes. This process, depicted in Fig. 2A, unfolded through a sequence of sequential steps, delineated below for clarity:

Initially, we accessed the Nobel Prize papers from the MAG database, a pivotal repository of scholarly endeavor. We extracted pertinent details, including the year of publication, title, journal, volume, issue, and the subject classification number. This trove of information was harnessed to pinpoint the specific domains within which the Nobel Prize-winning papers found their scholarly habitat.

Subsequently, we identified papers within the MAG database that shared congruent publication timelines, issues, and volumes with the Nobel Prize-winning counterparts. These identified papers were meticulously sifted to ascertain alignment with the same subject category, as determined by the OECD second-level classification number, serving as the bedrock for the assembly of the control paper set. It is pertinent to note that, within the scholarly milieu, journals stand as potent determinants of disciplinary categorization (86), and the second-level classification framework devised by the Organization for Economic Co-operation and Development (OECD) partially addresses the intricate challenge of interdisciplinary research classification, particularly pertinent for comprehensive journals.

The instantiation of the control group occupies a pivotal niche within our study, as it enables robust comparisons between the citation trajectories of Nobel Prize-winning papers and their analogs within the same academic arenas. This control group's efficacy hinges on the methodological construct that ensures the pairing of papers published contemporaneously, within identical journals, and encapsulating the same thematic domain. By adhering to this methodological rigor, the potential for confounding variables that might impinge upon the results is notably minimized.

**3.3 Disruptive Citation (DC)**

The CD index (22, 24) is expressed through the formula: $CD\ index = \frac{DC-CC}{DC+CC+RC}$, wherein R, FP, and C collectively constitute a cohesive knowledge flow sub-network, serving as a conduit for profound insights into the citation dynamics of pivotal nodes within the citation network. This index serves as a quantifier of the degree of either consolidation or disruption within the citation pattern of a given focal paper, with its quantification ranging from -1 (signifying complete consolidation) to +1 (indicating total disruption).

Building upon the foundation laid by our prior endeavors (8, 26), we present a pioneering methodology that introduces a novel dimension into citation analysis. This innovative approach delineates the bifurcation of citations associated with a central publication into two distinctive categories: disruptive and consolidating citations. Concretely, within the context of the citation network, the focal paper (FP) serves as the epicenter, accompanied by its ensemble of references, denoted as $R = \{r_1, r_2, \cdots, r_m\}$, and the collection of citations it has accrued, designated as $C = \{c_1, c_2, \cdots, c_n\}$. Together, these components coalesce into an interconnected sub-network that encapsulates the dynamics of knowledge dissemination, elucidating both forward and backward pivotal nodes within the citation ecosystem. In a quest to unearth the implicit interconnections pervading articles, we extend our purview to encompass the citation set of $R$, aptly labeled $RC = \{rc_1, rc_2, \cdots, rc_k\}$. It's pertinent to acknowledge that nodes within RC might encompass a broader array, emanating from their propensity to point towards multiple articles. With this nuanced foundation, the construct of consolidating citation (CC) emerges as the intersection between RC and C. Concomitantly, the disruptive citation (DC) unfolds as the discernible distinction, represented by DC=C-CC. This bifurcation delineates a fresh avenue for comprehending the impact of a given paper. By encompassing both the disruptive and consolidating facets inherent within its citations, this approach engenders a holistic comprehension of the paper's influence upon the research landscape. Beyond mere assessment, this approach resonates profoundly with its potential to gauge the paper's propensity for fostering disruptive innovation, thus propelling the boundaries of scholarly exploration and transformative endeavors.

While it is true that DC is a component of citation (C), we would like to emphasize that the novelty of DC lies in its specific focus on the disruptive aspect of impact. DC not only considers the overall impact represented by C but also incorporates the degree of disruption caused by a citing paper. This allows for a more nuanced assessment of the disruptive impact of scientific breakthroughs. In contrast, metrics like the CD-index use the normalization process to help reverse the direction of the correlation and provides a measure of disruption that is independent of the overall impact. However, the normalization method has raised many issues, such as the limitation of the CD index in capturing the true disruptive impact and its lack of convergent validity in identifying milestones (72-74, 87). We argue that while citation can capture the overall impact of papers, the CD index primarily reflects the level of disruption rather than the disruptive impact itself. DC, on the other hand, offers a more direct and comprehensive measure of disruptive impact by taking into account both the impact and degree of disruption. This unique perspective allows researchers to gain a deeper understanding of the transformative nature of scientific breakthroughs. Ultimately, this innovative methodology advances our comprehension of the complex interplay between scientific breakthroughs, their impact on the research community, and their potential to shape the course of future scientific inquiry.

Scientific breakthroughs inherently wield the prowess to upset and recalibrate established research paradigms within the complex labyrinth of scientific knowledge (88, 89). In so doing, they unveil novel corridors for exploration and inquiry. DC serves as an emblem of this profound dynamic, offering an encompassing grasp of the labyrinthine knowledge exchange within the citation network (20). Through its lens, we gain the capacity to differentially distill the impact of disruptive forces and consolidating influences amid the fabric of scientific literature. By weaving together the dimensions of disruptiveness and impact intensity, the Disruptive Citation (DC) introduces a finely-tuned metric, offering a nuanced gauge of the extent to which a pivotal paper instigates a seismic shift or fortifies the existing research paradigm. Beyond mere metrics, DC fathoms the underlying currents shaping the voyage of scientific progress, enriching our comprehension of the multi-faceted panorama of knowledge creation and its diffusion.

## 4 Results

### 4.1 Descriptive analysis

As illustrated in Fig. 1B-E, both the number of Nobel Prize-winning papers and the control group are uniformly distributed across the three fields, and display a highly similar distribution in terms of publication year. It was somewhat unexpected to observe that the citations of papers in both samples show a gradual increase over time, despite the widely held belief in the first-mover advantage (90). This trend is related to the concept of citation inflation (91), which has been widely discussed in recent years, and could be attributed to retrospective and prospective reasons (92), such as the steadily growing of the total production of citations (91), the aging effect (93), the decay of collective memory and attention (94), or the fact that citation patterns before 1950 were fundamentally different from those that followed (95).

It is worth noting that we did not set citation windows in our study, as the majority of our sample data (99%) were published before 2000. This implies that all papers in the sample have a citation window of at least 20 years. In addition to ensuring a consistent and comparable dataset, the approach we adopted also ensures that papers published in earlier periods, such as the 19th century, receive adequate citations. This is because our study did not set citation windows, which allowed for a more extended period of time for papers to accumulate citations. As a result, we were able to include a diverse range of papers, regardless of their publication year, in our investigation. This is an important aspect of our analysis, as it allows us to gain insights into the long-term impact of scientific breakthroughs, which may take several decades or even centuries to become widely recognized and appreciated (96).

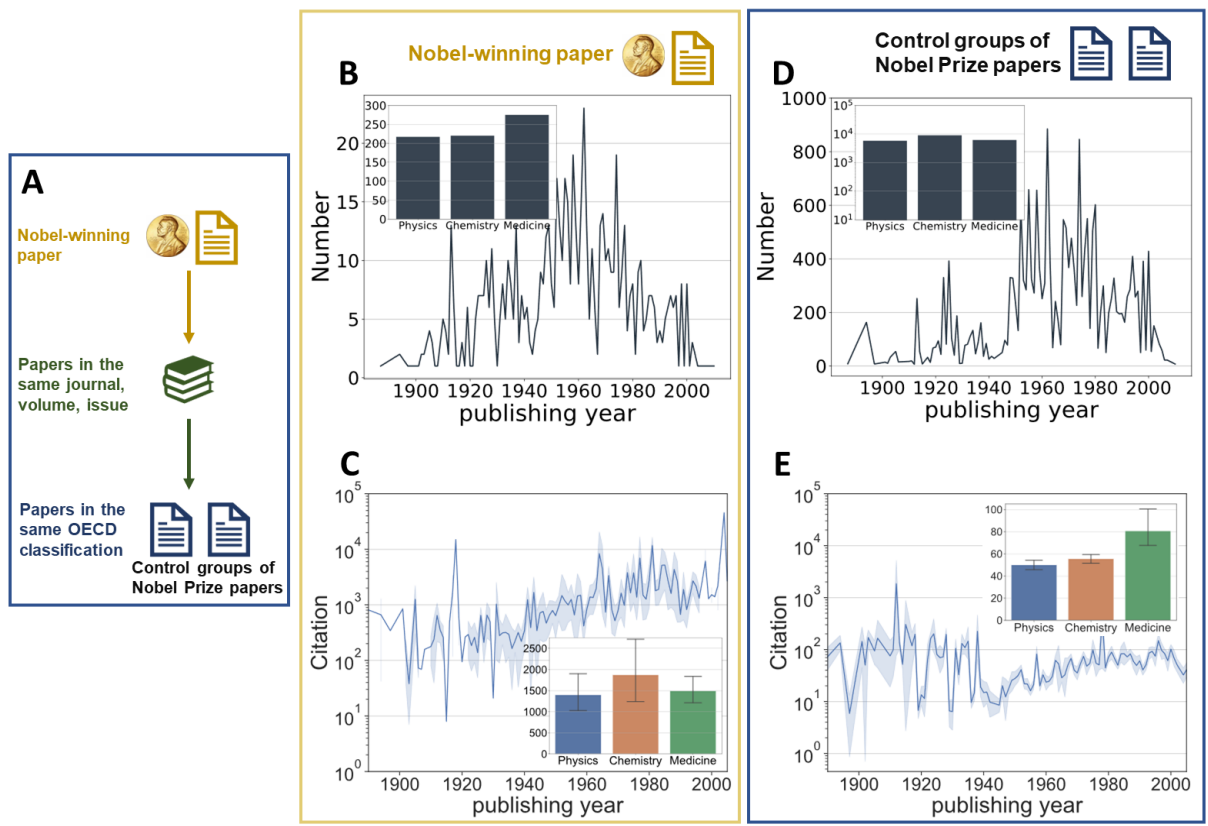

**Fig. 1. Nobel-Winning papers and control groups: procedures and descriptive statistics. (A)** the process for generating control groups of Nobel-winning papers in the MAG dataset. **(B-C)** descriptive statistics and distributions of the number and average citation of Nobel-winning papers over time and across fields. **(D-E)** descriptive statistics and distributions of the number and average citation of control group papers over time and across fields.

### 4.2 Comprehensive Comparative Analysis

To inaugurate our exploration, we embark upon an exhaustive scrutiny of Nobel milestone papers alongside their corollary control cohorts. This comprehensive assessment is underpinned by the dynamic interplay of Disruptive Citation (DC), CD-index, and supplementary disruptive indicators, within the three distinct spheres of Physics, Chemistry, and Medicine. In this endeavor, we weave a compelling narrative that unravels the transformative ripples echoing from Nobel-winning papers, accentuating the stark differentials that demarcate these seminal works from their associated control assemblages.

As artfully depicted in Fig. 2A, Nobel laureate papers across the triad of disciplines distinctly unveil a notably amplified Disruptive Citation (DC) quotient when juxtaposed with their analogous control groups. This discernible divergence in DC stands at 96.9%, 97.6%, and 96.1% for Physics, Chemistry, and Medicine, respectively. This substantive discrepancy underscores the discernible propensity of Nobel-winning papers to wield an elevated degree of disruptive resonance vis-à-vis their peers, gauged through the prism of Disruptive Citation. Importantly, this disparity in DC's magnitude is underscored by the two-tailed Mann-Whitney U-test, which accentuates the statistical significance of these deviations, as reflected by p-values clocking in below the remarkable threshold of $10^{-100}$ across all three disciplines.

Contrastingly, the CD-index—depicted in Fig. 2B—portrays a more marginal disparity between Nobel-winning papers and their control ensembles. To be precise, the CD-index of Nobel-winning papers tallies at 0.74, 0.53, and 0.41 for Physics, Chemistry, and Medicine, in that order. Correspondingly, their control group counterparts exhibit CD-index scores of 0.66, 0.48, and 0.36. The ensuing discrepancy stands at 9.8%, 9.7%, and 9.6% for Physics, Chemistry, and Medicine, respectively. While the disparities bear statistical significance in Medicine ($p < 10^{-6}$), this significance is variably observed in Physics ($p = 0.017$) and Chemistry ($p = 0.003$) at the 0.001 threshold.

The findings derived from our rigorous analytical expedition duly pronounce that DC emerges as a more potent and adept metric in assessing the disruptive impact of papers, as juxtaposed against the conventional CD-index. The robustness of this assertion arises from the salient distinction that DC meticulously accounts for the extent of disruption inflicted upon the citation network by the citing paper, a dimension that stands conspicuously absent within the purview of CD-index, which predominantly accords significance to the chronological trajectory of citations. As the curtains draw on our investigation, the resounding conclusion that surfaces is the imperative mandate to incorporate DC as a paramount metric, wielding unparalleled efficacy in the appraisal of the reverberations unleashed by scientific breakthroughs.

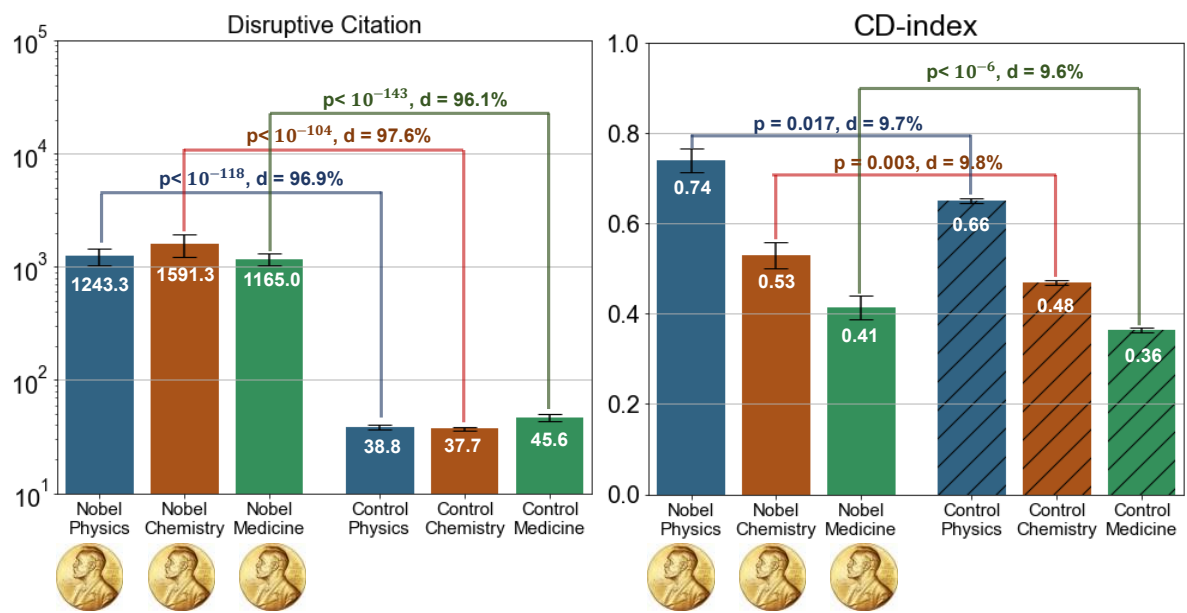

**Fig. 2. Comparison of Nobel-winning papers and control groups: Disruptive Citation (DC) vs. CD-Index (A)** The Nobel-winning papers have significant larger Disruptive Citation (DC) than control groups. **(B)** The Nobel-winning papers have slightly larger CD-index than control groups.

To fortify the tenets of our analysis, a comprehensive probe was extended across the entirety of our sample dataset, encompassing both the echelons of Nobel-winning papers and their accompanying control groups. Within this expansive purview, we judiciously culled the top five papers in each field boasting the loftiest Disruptive Citation (DC) scores. Significantly, this elite cohort was unanimously composed of Nobel-winning papers, unequivocally attesting to their paradigm-shifting influence within the scientific tapestry. The tableau elucidated in Table 2 unveils a striking panorama: in the realm of Physics, the paper published in 2004 within the annals of Science reigns supreme with a towering DC of 44,083 and a CD-index of 0.70; within Chemistry, the laureate paper emanating from PNAS in 1977 commands attention with a formidable DC of 55,018 and a CD-index of 0.81. Similarly, the domain of Medicine finds resonance in the paper published in 1981 within the pages of *Pflügers Archiv: European Journal of Physiology*, boasting a DC of 15,136 alongside a CD-index of 0.68.

Curiously, we unearthed the paradoxical revelation that select paramount papers within each domain manifested conspicuously low CD-index scores. Strikingly, a significant proportion of papers, spanning Nobel-winning gems and control groups alike, exhibited a CD-index of 1, rendering the distinction between these groups a challenging endeavor based solely on this metric. This intriguing revelation underscores the limitations of the CD-index in distinguishing the profound impact of Nobel-winning papers vis-à-vis their control cohorts. This phenomenon underscores the pivotal role carved by Disruptive Citation (DC) as a poignant metric for gauging the transformative repercussions of papers.

**Table 2. The top papers with the highest Disruptive Citation in each field.**

| Top 5 papers with the highest Disruptive Citation (DC)-Physics | | | | | | | |
|---|---|---|---|---|---|---|---|
| Title | Year | Journal | Ref | Cit | DC | CD-index | Type |
| ELECTRIC FIELD EFFECT IN ATOMICALLY THIN CARBON FILMS | 2004 | Science | 13 | 45519 | 44083 | 0.70 | Nobel |
| OBSERVATIONAL EVIDENCE FROM SUPERNOVAE FOR AN ACCELERATING UNIVERSE AND A COSMOLOGICAL CONSTANT | 1998 | The Astronomical Journal | 117 | 14791 | 9720 | 0.12 | Nobel |
| POSSIBLE HIGH TC SUPERCONDUCTIVITY IN THE BA LA CU O SYSTEM | 1986 | European Physical Journal B | 19 | 9365 | 8938 | 0.75 | Nobel |
| THEORY OF SUPERCONDUCTIVITY | 1957 | Physical Review | 0 | 8437 | 8437 | 1.00 | Nobel |
| ABSENCE OF DIFFUSION IN CERTAIN RANDOM LATTICES | 1958 | Physical Review | 0 | 8036 | 8036 | 1.00 | Nobel |
| GIANT MAGNETORESISTANCE OF 001 FE 001 CR MAGNETIC SUPERLATTICES | 1988 | Physical Review Letters | 0 | 6833 | 6833 | 1.00 | Nobel |
| **Top 5 papers with the highest Disruptive Citation (DC)-Chemistry** | | | | | | | |
| Title | Year | Journal | Ref | Cit | DC | CD-index | Type |
| DNA SEQUENCING WITH CHAIN TERMINATING INHIBITORS | 1977 | PNAS | 12 | 57449 | 55018 | 0.81 | Nobel |
| SELF CONSISTENT EQUATIONS INCLUDING EXCHANGE AND CORRELATION EFFECTS | 1965 | Physical Review | 0 | 38949 | 38949 | 1.00 | Nobel |
| INHOMOGENEOUS ELECTRON GAS | 1964 | Physical Review | 0 | 32860 | 32860 | 1.00 | Nobel |
| THE ADSORPTION OF GASES ON PLANE SURFACES OF GLASS MICA AND PLATINUM | 1918 | Journal of the American Chemical Society | 0 | 14924 | 14924 | 1.00 | Nobel |
| C 60 BUCKMINSTERFULLERENE | 1985 | Nature | 6 | 11964 | 11414 | 0.80 | Nobel |
| PRIMER DIRECTED ENZYMATIC AMPLIFICATION OF DNA WITH A THERMOSTABLE DNA POLYMERASE | 1988 | Science | 19 | 15232 | 8679 | 0.01 | Nobel |
| **Top 5 papers with the highest Disruptive Citation (DC)-Medicine** | | | | | | | |
| Title | Year | Journal | Ref | Cit | DC | CD-index | Type |
| IMPROVED PATCH-CLAMP TECHNIQUES FOR HIGH-RESOLUTION CURRENT RECORDING FROM CELLS AND CELL-FREE MEMBRANE PATCHES | 1981 | Pflügers Archiv: European Journal of Physiology | 22 | 16411 | 15136 | 0.68 | Nobel |
| A QUANTITATIVE DESCRIPTION OF MEMBRANE CURRENT AND ITS APPLICATION TO CONDUCTION AND EXCITATION IN NERVE | 1952 | The Journal of Physiology | 12 | 17009 | 15002 | 0.60 | Nobel |
| CONTINUOUS CULTURES OF FUSED CELLS SECRETING ANTIBODY OF PREDEFINED SPECIFICITY | 1975 | Nature | 14 | 14594 | 13910 | 0.65 | Nobel |
| A COMPREHENSIVE SET OF SEQUENCE ANALYSIS PROGRAMS FOR THE VAX | 1984 | Nucleic Acids Research | 10 | 13289 | 12523 | 0.44 | Nobel |
| THE GENETICS OF CAENORHABDITIS ELEGANS | 1974 | Genetics | 3 | 12240 | 11949 | 0.92 | Nobel |
| INDUCTION OF PLURIPOTENT STEM CELLS FROM MOUSE EMBRYONIC AND ADULT FIBROBLAST CULTURES BY DEFINED FACTORS | 2006 | Cell | 50 | 19076 | 11743 | 0.08 | Nobel |

### 4.3 Empirical Validation and Regression Analysis

The validation of Disruptive Citation (DC) as a robust metric beckoned an intricate regression analysis, designed to unravel the predictive efficacy of Nobel-winning papers, deftly codified as binary variables, vis-à-vis the DC landscape. This meticulous exploration was conducted against the backdrop of meticulous control for confounding variables, including citations, reference counts, team dimensions, page lengths, and the strategic incorporation of journal fixed effects and year fixed effects.

Given the inherent skewness endemic to the DC distribution, the regression apparatus opted for a strategic embrace of the Negative Binomial model, as elucidated by Davies, Gush (97). The culmination of this empirical voyage unfurls through Table 3, where the substantive results are disseminated. Of paramount significance is the coefficient governing the Nobel-winning papers, steadfastly yielding a consistent trajectory across the four sequential iterations of the regression framework (1-4). The lofty significance and positive orientation of this coefficient epitomize the tangible imprint of Nobel-winning papers upon amplifying the DC quotient. This illumination persists even after the meticulous choreography of contending variables—citation, reference count, team magnitude, page expanse—alongside the judicious anchoring of journal and year fixed effects.

Moreover, the panorama unveiled by the regression tableau accentuates the commanding roles executed by the reference count and team dimensions. A conspicuous alignment toward significant and positive coefficients coalesces, wherein elevated reference counts and expansive team ensembles lay the foundation for augmented DC values. Conversely, the realm of page length betrays a lack of significance, steadfastly averring that the augmentation of page expanse does not wield an appreciable influence upon DC dynamics.

Furthermore, the ambit of the regression enigma extends its embrace to the CD-index, dissected through models 5-8. Evidently, the mantel of Nobel laureates casts a discernible imprint upon the CD-index realm as well, albeit with a semblance of moderation. This underscores the assertion that while the CD-index does capture the incipient echoes of Nobel-winning papers, its potency pales in comparison to the pivotal role mastered by the DC metric. A conspicuous divergence emerges in the realm of DC and CD-index, delineating the more refined utility and discernment offered by DC in decoding the disruptive resonance sewn into the fabric of Nobel-winning papers.

**Table 3. Regression analysis results.**

| Dependent Variable | Disruptive Citation (DC) | | | | CD-Index | | | |
|---|---|---|---|---|---|---|---|---|
| Regression Model | Negative Binomial | | | | OLS | | | |
| | (1) | (2) | (3) | (4) | (5) | (6) | (7) | (8) |
| Nobel=1 | 3.9651*** (0.041) | 3.8513*** (0.0412) | 0.7671*** (0.0403) | 1.1358*** (0.0435) | 0.0511*** (0.0146) | 0.107*** (0.0136) | 0.1115*** (0.0172) | 0.0885*** (0.0143) |
| Citation | | | 0.0039*** (0.0) | 0.0034*** (0.0) | | | 0.0*** (0.0) | 0.0*** (0.0) |
| #Reference | | 0.0356*** (0.0008) | -0.0042*** (0.0006) | 0.013*** (0.0008) | | -0.0141*** (0.0003) | -0.0219*** (0.0003) | -0.0142*** (0.0003) |
| Team Size | | 0.0311*** (0.0009) | 0.0554*** (0.0008) | 0.0076*** (0.0009) | | 0.0001 (0.0003) | 0.0022*** (0.0004) | 0.0001 (0.0003) |
| Page Length | | 0.0001 (0.0) | -0.0001 (0.0001) | -0.0001 (0.0) | | 0.0 (0.0) | 0.0* (0.0) | 0.0 (0.0) |
| Intercept | 2.7282*** (0.0081) | 1.9575*** (0.4086) | 2.6321*** (0.4085) | 2.0692*** (0.4119) | 0.0862 (0.1422) | 0.3142** (0.1343) | 0.612*** (0.0034) | 0.3172** (0.1343) |
| Journal Fixed Effect | Yes | Yes | No | Yes | Yes | Yes | No | Yes |
| Year Fixed Effect | Yes | Yes | No | Yes | Yes | Yes | No | Yes |
| Observations | 21666 | 21666 | 21666 | 21666 | 21666 | 21666 | 21666 | 21666 |
| Adjusted $R^2$ | - | - | - | - | 0.469 | 0.547 | 0.265 | 0.547 |
| Pseudo $R^2$ | 0.8257 | 0.8424 | 0.8872 | 0.9148 | - | - | - | - |

Note: *** $p<0.001$, ** $p<0.01$, * $p<0.05$. The coefficients and their corresponding robust standard errors are reported for each variable in each model.

**4.4 Team size and disruptive impact**

To delve deeper into the intricate interplay between team size and the disruptive impact of scientific endeavors, we draw upon the seminal work of Wu, Wang (23), who illuminated the intricate dance between large teams fostering scientific development and small teams heralding disruptive potential. Within this analytical framework, we leverage our expansive sample dataset, embracing the entirety of papers published in the same journal and year as the curated cohort of 712 Nobel-winning papers spanning the triptych of disciplines.

Our inquiry unfurls through a two-fold prism. First, we embark on a direct comparison of Disruptive Citation (DC) metrics across a spectrum of team sizes. This panoramic exploration not only sheds light on the variegated dimensions of disruptive impact across diverse team compositions but also enables us to decipher the nuanced relationship between team size and disruptive resonance. Second, our analysis adopts a stratified partitioning of the sample dataset within each field, demarcating the bounds of disruption and disruptive impact as appraised by the CD-index and DC, respectively. The ensuing dissection meticulously appraises the divergence in team sizes embedded within these distinct subsets. This judicious bifurcation allows us to meticulously probe the potential correlation between team size and the degrees of disruption and disruptive impact inherent within a given field.

The unveiled findings, cast into the canvas of Fig. 3, cast a revealing light on the landscape of team size and its orchestration with disruptive impact. In this intricate tapestry, we observe a nuanced dance wherein large teams tend to manifest marginally diminished CD-index values, with a singular exception in the realm of single-author papers (team size = 1). However, the prevailing disparities in team size when tethered to high and low disruption thresholds, as defined by the CD-index metric, remain evasive and inconspicuous (Fig. 3C).

In stark contradistinction, a resounding symphony reverberates when unraveling the nexus between team size and disruptive impact (DC). Across all three disciplines, a conspicuous phenomenon, accentuating the propensity of larger teams to wield a more potent disruptive impact, as resonated by higher DC values. Moreover, the symposium of analysis unveils a resolute reality—papers epitomizing elevated disruptive impact invariably bear the hallmark of substantially augmented team dimensions. This symphony amplifies the association between team size and disruptive impact, underscoring that the monumental quest to disrupt the existing research paradigm finds its ally in the realms of extensive collaboration, profound expertise, and ample resources—the signature assets of larger teams.

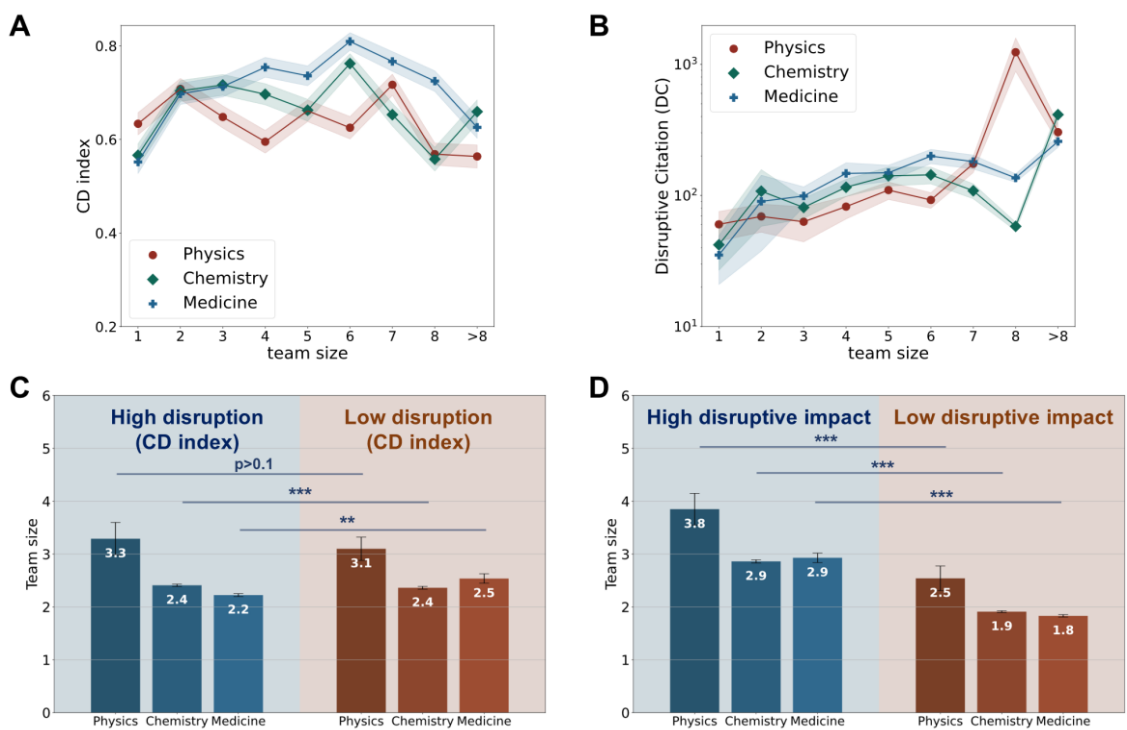

**Fig. 3: Relationship between Team Size and Disruption (CD-index), and Team Size and Disruptive Impact (DC). (A-B)** The change in CD-index and DC values across different team sizes, accompanied by bootstrapped 95% confidence intervals to provide a measure of statistical uncertainty. **(C)** The differences in team size between papers with high levels of disruption (based on CD-index) and those with low levels of disruption. **(D)** The variations in team size between papers with high and low disruptive impact. Significance levels are denoted as *** for $p < 0.001$ and ** for $p < 0.01$, by Mann-Whitney U test.

### 4.5 Linguistic feature and disruptive impact

In pursuit of unraveling the intricate connection between linguistic attributes and the variegated tapestry of disruptive impact, we embarked on a meticulous linguistic analysis, poised to discern the distinctive linguistic hallmarks underpinning papers that traverse the spectrum of disruptive impact. The trajectory of our inquiry, as chronicled through the illustrative vistas of Fig. 4, bears witness to the compelling presence of Nobel-winning papers within the echelons of high disruptive impact—a testament to the potency of this linguistic attribute differential in encapsulating the essence of scientific breakthroughs. Of particular note, the revelation of an expanse in title length disparities between papers harboring high and low disruptive impact stands as a towering monolith among our findings. Papers that emerged as flagbearers of high disruptive impact unveiled titles of considerable length, casting a striking juxtaposition against their low disruptive impact counterparts.

Venturing further into the chasms of linguistic nuances that pervade the realm of high disruptive impact papers, our investigation peeled back the layers of word utilization patterns. This expedition bore forth a fascinating revelation—a propensity among high disruptive impact papers to embrace a copious array of verbs and nouns steeped in the lexicon of specialized jargon and technical terminology. The tapestry of verbs woven into their narrative included "catalyze" and "clone," while the fabric of nouns enfolded terms such as "cell" and "DNA." These findings bear eloquent testimony to the inclination of high disruptive impact papers to wield the armory of technical parlance, thus mirroring the erudite and innovative essence of the research they encapsulate.

Conversely, the portals of our inquiry also widened to encompass the delineation of disparities between papers adorned with high and low CD-index, as the tableau depicted in Fig. 5 so vividly portrays. Astonishingly, our foray into this terrain unearthed a revelation—papers espousing high disruption, as etched by the CD-index, failed to exhibit a conspicuous influx of Nobel-winning papers, contrasting the narrative that unfolded in the realm of high disruptive impact. Moreover, the troves of title length and lexical variances between papers straddling the dichotomy of high and low disruption appeared enshrouded in a shroud of insignificance, echoing a resounding note that CD-index may not emerge as a beacon to effectively sift scientific breakthroughs from their non-disruptive counterparts. This absence of differentiation underscores the incapacity of CD-index to encompass the intricate dimensions associated with disruptive impact.

These revelations, like gleaming gems unearthed from the heart of an academic mine, resound with a clear clarion call for the ascendancy of Disruptive Citation (DC) as a metrical touchstone, a beacon to illumine the terrain of disruptive impact in scientific discourse. The linguistic tapestry we wove through the prism of high disruptive impact papers, replete with lengthier titles and an immersion in the sea of technical parlance, casts the luminous spotlight on the innovative and transformative vista that these papers embody. DC, bearing the imprimatur of holistic measurement, not only captures the disruptive force of scientific breakthroughs but also offers glimpses into the linguistic nuances that underscore such impact—a pantheon that deepens our comprehension of the intricate ballet that defines disruptive research.

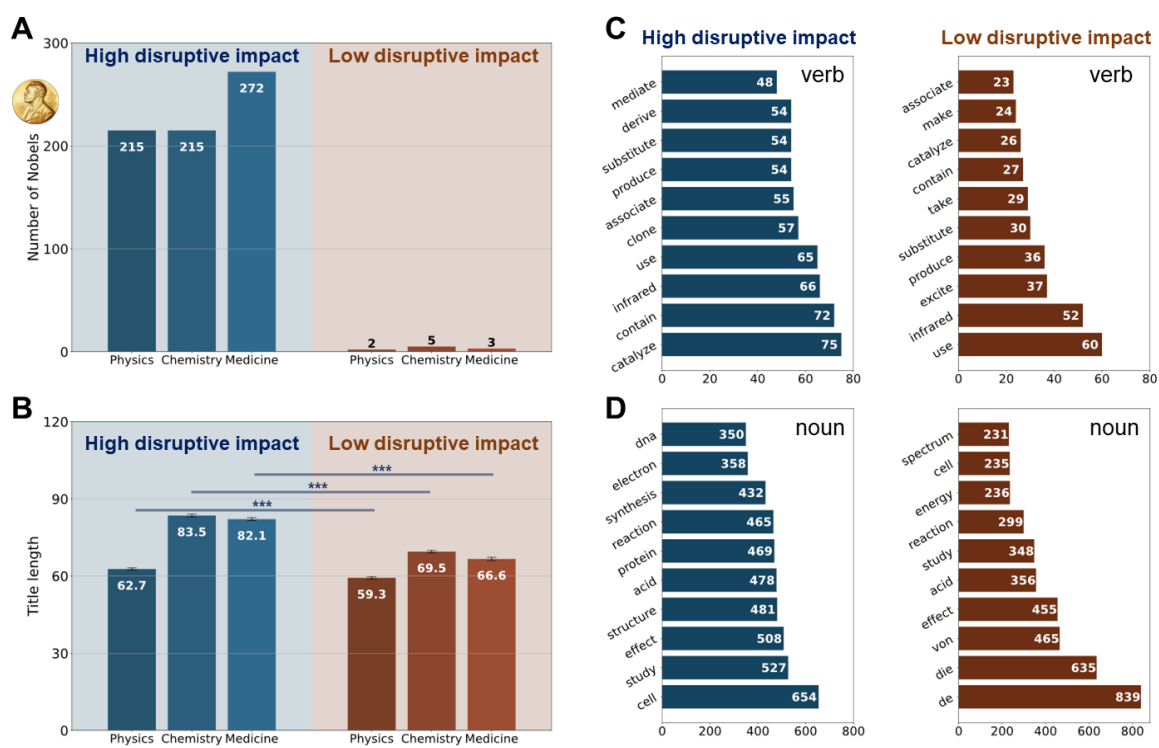

**Fig. 4. Linguistic differences for high disruptive impact papers and low disruptive impact ones. (A-B)** The variations in number of **(A)** Nobel-winning papers and **(B)** title length between papers with high levels of disruptive impact (based on DC) and those with low levels of disruptive impact. Significance levels are denoted as *** for $p < 0.001$ and ** for $p < 0.01$, by Mann-Whitney U test. **(C-D)** Frequency of occurrence of verbs and nouns in titles and abstract (if possible) for papers with high and low disruptive impact.

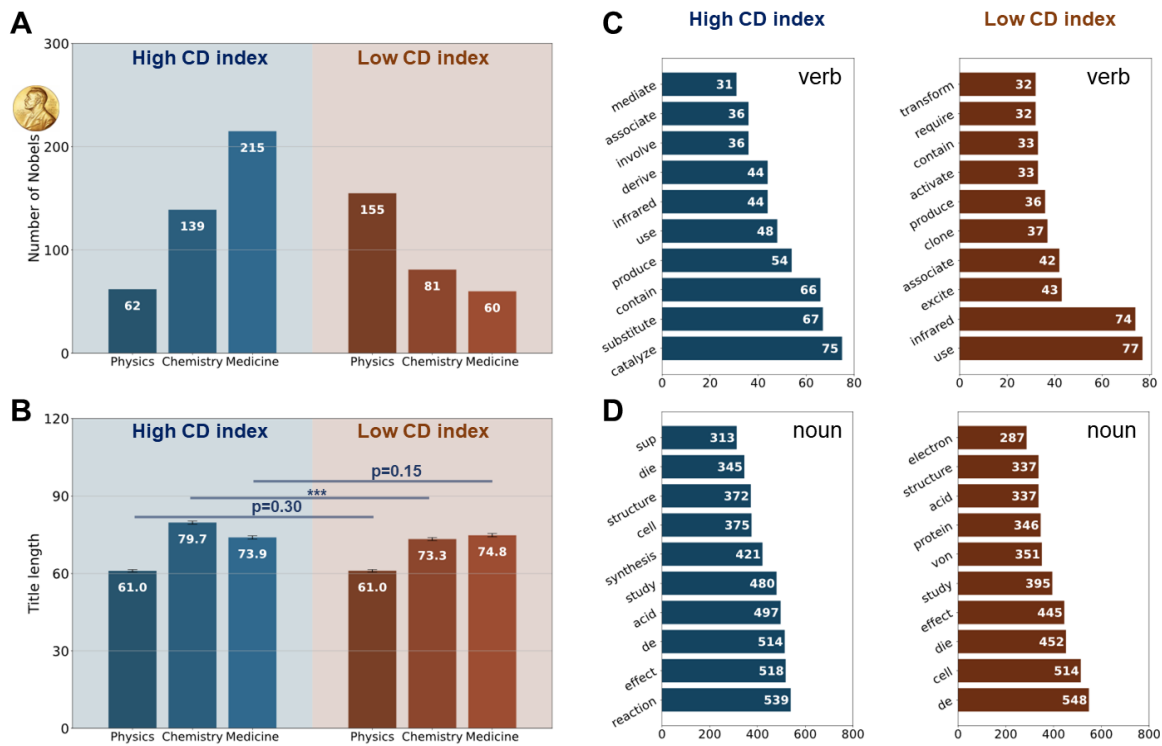

**Fig. 5. Linguistic differences for high and low CD-index. (A-B)** The variations in number of (**A**) Nobel-winning papers and (**B**) title length between papers with high levels of disruptive impact (based on DC) and those with low levels of CD-index. Significance levels are denoted as *** for p < 0.001 and ** for p < 0.01, by Mann-Whitney U test. (**C-D**) Frequency of occurrence of verbs and nouns in titles and abstracts (if possible) for high disruptive impact papers vs low disruptive impact papers.

## 5 Discussion

### 5.1 Theoretical Implications

The findings of this study carry several important theoretical implications for the field of scientific research evaluation and the understanding of disruptive impact. First, the research illuminates the limitations of conventional metrics such as the CD-index in capturing the multifaceted nature of disruptive impact. By introducing DC, which incorporates both the disruptive influence and overall impact, this study contributes to the development of more comprehensive and refined metrics for assessing the groundbreaking influence of scientific papers. This advancement addresses a significant gap in impact measurement techniques, enabling a more nuanced evaluation of the true transformative power of research contributions. By focusing on the disruptive aspect of impact, DC enriches our understanding of the intricate relationships between papers, revealing the extent to which citing works deviate from prevailing citation patterns. This insight into the disruptive potential of papers contributes to a more holistic comprehension of the profound contributions that reshaped the scientific landscape. The integration of DC offers novel insights into the dynamics of knowledge flow within the scientific community. By exploring the intricate connections between citing and cited papers, this research uncovers the disruptive threads that interlace scientific paradigms. This deeper understanding of how scientific breakthroughs challenge, modify, or reinforce established knowledge structures advances our comprehension of the evolving trajectories of research disciplines.

Second, the study contributes to the theoretical discourse on the relationship between team size and disruptive impact (23, 98). By examining the differences in disruptive impact based on team size, our findings argue previous research suggesting that larger teams tend to generate more consolidating knowledge, while smaller teams have the potential to disrupt the scientific landscape (23). The observation that larger teams exhibit higher disruptive impact as measured by DC provides empirical evidence supporting the notion that collaboration plays a significant role in scientific breakthroughs. These findings contribute to the understanding of the complex interplay between team characteristics and disruptive impact, shedding light on the factors that drive scientific innovation.

Furthermore, the analysis of linguistic features associated with disruptive impact adds to the theoretical understanding of how language reflects and influences scientific breakthroughs (23, 24). The finding that papers with high disruptive impact tend to have longer titles and employ more technical language suggests that the linguistic characteristics of scientific publications can serve as indicators of their transformative potential. This highlights the importance of language in conveying the innovative and disruptive nature of research and provides insights into the communication strategies employed by researchers to convey the significance of their work. These findings contribute to the theoretical discourse on the role of language style in scientific communication and the ways in which linguistic features can inform research evaluation and assessment.

**5.2 Practical Implications**

The findings of this study have important practical implications for various stakeholders involved in scientific research evaluation, funding allocation, and policymaking. The introduction of Disruptive Citation (DC) offers a valuable tool for assessing the disruptive impact of scientific papers in a more comprehensive and accurate manner. This metric can aid funding agencies, academic institutions, and policymakers in making informed decisions about resource allocation and the promotion of research excellence (99). This can help identify and support research that has the potential to challenge and reshape existing research paradigms (100), leading to future scientific advancements.

Moreover, the findings regarding the relationship between team size and disruptive impact have practical implications for research collaboration and team formation. The observation that larger teams tend to have higher disruptive impact suggests that fostering collaboration may be beneficial for generating groundbreaking discoveries. Academic institutions and research organizations can use this knowledge to promote collaborative initiatives and provide resources to support team-based research (97). Besides, Researchers can also use linguistic feature insights to enhance their communication strategies and ensure that their work is effectively conveyed to the scientific community and beyond. Science communicators can also leverage these findings to improve the dissemination of scientific breakthroughs to wider audiences, fostering a greater appreciation for the transformative potential of scientific research. By applying these insights, stakeholders in the scientific community can make informed decisions, allocate resources more effectively, and promote research that has the potential to drive societal progress and innovation.

**5.3 Limitation**

Despite the contributions of this study, it is essential to acknowledge its limitations. First, our analysis is restricted to the evaluation of Nobel-winning papers and their corresponding control groups, potentially limiting the generalizability of our findings. Future research could expand the scope of the analysis to include a more diverse range of scientific papers and disciplines, further validating the applicability of DC.

It is also important to note that DC, while providing a more comprehensive assessment of impact, is still a quantitative measure and may not capture the full complexity and context of scientific breakthroughs. The interpretation of disruptive impact is subjective to some extent, and there may be other qualitative factors that contribute to the evaluation of research excellence. Future studies could explore complementary qualitative approaches to further enhance the understanding of disruptive impact.

Lastly, this study focused on the analysis of team size and linguistic features as related to disruptive impact. While the findings provide valuable insights, there may be other factors and variables that influence disruptive impact, such as funding resources, institutional support, and individual researcher characteristics. Future research could consider a more comprehensive examination of these factors to provide a more holistic understanding of disruptive impact in scientific research.

**5.4 Outlook**

In light of the aforementioned limitations, several opportunities for future research arise. First, future investigations could extend the application of Disruptive Citation (DC) to additional scientific domains and evaluate the impact of a wider range of scientific papers. This would further validate the utility of DC and provide insights into the disruptive nature of scientific advancements across various fields.

Second, future research could explore the predictive power of DC, examining whether this metric can forecast the emergence of scientific breakthroughs and identify potential Nobel-winning research endeavors in their early stages. Such investigations could contribute to the development of novel research evaluation methodologies, benefiting policymakers and funding agencies in their decision-making processes.

Last, future studies could investigate the relationship between DC and other emerging indicators of research impact, such as Altmetrics (101). This line of inquiry would help to further refine the understanding of the multifaceted nature of scientific impact and provide valuable insights into the diverse ways in which scientific advancements influence and reshape the broader research landscape.

**Acknowledgments**

This research was supported by the Jiangsu Key Laboratory Fund, the International Joint Informatics Laboratory Fund, and the Fundamental Research Funds for the Central Universities.

**Conflict of Interest**

The authors declare no competing interest relevant to this article.

**CRediT authorship contribution statement**

**Alex J. Yang**: Writing-original draft, Conceptualization, Methodology, Software, Data visualization, Validation, Data curation, Resources. **Xiaohui Yan**: Writing- review & editing, Validation. **Yiqin Zhang**: Conceptualization, Writing- review & editing. **Hao Wang**: Writing- review & editing, Supervision. **Sanhong Deng**: Conceptualization, Methodology, Writing-review & editing, Supervision, Funding acquisition.

**Data Availability**

The Nobel-winning Papers data used in this study are open-access at Harvard Dataverse https://dataverse.harvard.edu/dataset.xhtml?persistentId=doi:10.7910/DVN/6NJ5RN. The MAG dataset can be accessed through the Get Microsoft Academic Graph on Azure storage - Microsoft Academic Services | Microsoft Learn. The version number of MAG data used in our study is the final (latest) version, published in December 20, 2021.

Other data used in this study can be obtained by making reasonable requests.

**Appendix**

**Note 1 Journals for Nobel-winning papers and control groups**

We present the most frequent journals in which our sample data were published. As shown in Fig. A1, our sample data were published in some of the most renowned and prestigious journals in each field, including PRL, Nature, Science, Cell, PNAS, and others. These journals are recognized as leaders in their respective fields and have a reputation for publishing high-quality research that makes significant contributions to the scientific community. This also ensures that the papers included in our investigation have undergone rigorous peer-review processes and have been deemed to be of high scientific quality and significance. The similarity in the distribution of Nobel Prize-winning papers and the control group across different fields and publication years strengthens the validity of our investigation. The absence of any significant differences in these factors between the two samples helps to ensure that any differences observed in the citation patterns of disruptive and consolidating citing papers are not influenced by extraneous factors.

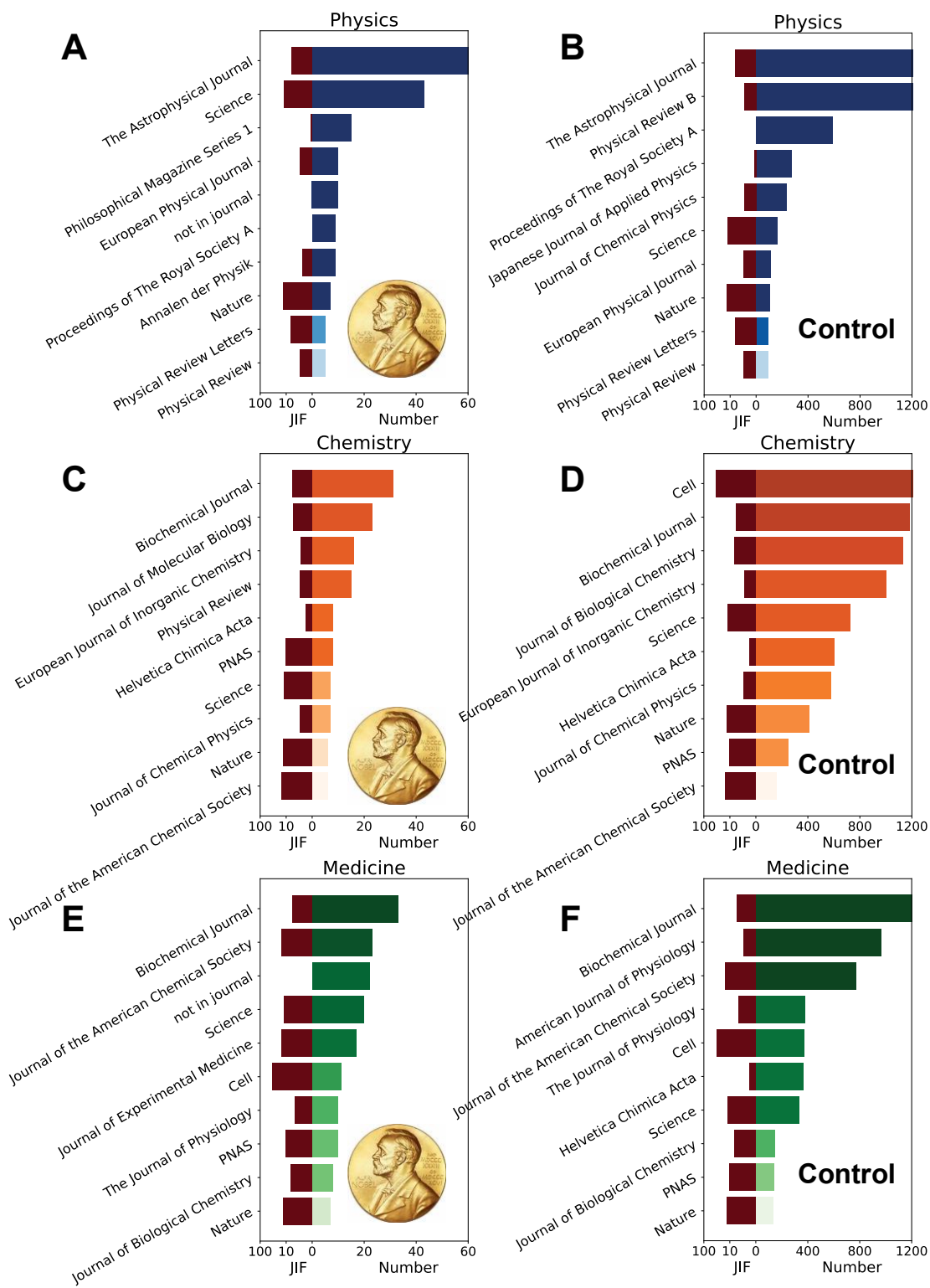

**Fig. A1. Impact factor and number of top-10 journals for Nobel-winning papers and control groups in (A-B) Physics, (C-D) Chemistry, and (E-F) Medicine.** The impact factor (JIF) is calculated in the Microsoft Academic Graph (MAG) dataset.

**Note 2 Alternative scientific breakthroughs**

While the use of Nobel-winning papers as a reflection of scientific breakthroughs is customary, we also offer an alternative way with extant efforts. Wu et al. (23) conducted a survey in which respondents had been asked to nominate "disrupting papers". We matched these papers in MAG dataset, and calculate their DC and CD-index, as shown in Table A1. We show that these alternative scientific breakthroughs all get very high Disruptive Citation (DC), yet their CD-index can be relatively low (0.2).

However, it is important to note that the samples of scientific breakthroughs identified through this survey are relatively small in number, and not all of them can be matched in the Microsoft Academic Graph (MAG) dataset. Additionally, many of the papers identified through the survey are also Nobel-winning papers. As a result, we have focused our analysis on Nobel-winning papers, which are widely recognized as significant scientific breakthroughs and serve as a customary benchmark in the field.

**Table A1 Selected scientific breakthroughs nominated by survey in Wu, Wang (23).**

| Contribution | MAGid | Author | Year | Journal | Title | Disruption | DC |
|---|---|---|---|---|---|---|---|
| Info-metrics | 2088209891 | Price, D. J. D. S. | 1965 | Science | Networks of scientific papers. | 0.92 | 2082 |
| Density functional Theory | 2030976617 | Hohenberg, P., & Kohn, W. | 1964 | Physical Review | Inhomogeneous electron gas. | 0.87 | 32860 |
| Nonlinear Dynamics | 2141394518 | Lorenz, E. N. | 1963 | Journal of the atmospheric sciences | Deterministic nonperiodic flow. | 0.86 | 13215 |
| Networks | 2112090702 | Watts, D. J., & Strogatz, S. H. | 1998 | Nature | Collective dynamics of small-world networks | 0.48 | 26028 |
| Latent Dirichlet Allocation Model | 1880262756 | Blei, D. M., Ng, A. Y., & Jordan, M. I. | 2003 | Journal of machine Learning research | Latent dirichlet allocation | 0.43 | 19253 |
| Rubin Causal Model | 2150291618 | Rosenbaum, P. R., & Rubin, D. B. | 1983 | Biometrika | The central role of the propensity score in observational studies for causal effects. | 0.43 | 16201 |
| The Steady State Theory of Universe | 2076480526 | Hoyle, F. | 1948 | Monthly Notices of the Royal Astronomical Society | A new model for the expanding universe | 0.42 | 459 |
| Infants' and Children's Early Numerical Cognition | 2067253250 | Wynn, K. | 1992 | Nature | Addition and subtraction by human infants. | 0.36 | 585 |
| LASSO Regression | 2135046866 | Tibshirani, R. | 1996 | Journal of the Royal Statistical Society | Regression shrinkage and selection via the lasso. | 0.32 | 26049 |
| Game Theory | 2330024298 | Nash, J. | 1951 | Annals of mathematics | Non-cooperative games. | 0.28 | 3821 |
| Yang–Mills Theory | 2232347689 | Yang, C. N., & Mills, R. L. | 1954 | Physical Review | Conservation of isotopic spin and isotopic gauge invariance. | 0.27 | 2164 |
| Metabolic theory of Ecology | 2080843536 | West, G. B., Brown, J. H., & Enquist, B. J. | 1997 | Science | A general model for the origin of allometric scaling laws in biology. | 0.27 | 2387 |
| Simulated Annealing | 2024060531 | Kirkpatrick, S., Gelatt, C. D., & Vecchi, M. P. | 1983 | Science | Optimization by simulated annealing. | 0.2 | 27099 |

**Note 3 Benchmark measures**

We used several benchmark measures to demonstrate the effectiveness of DC in reflecting a paper's disruptive impact. The selected benchmark measures include those that measure the impact of papers and those that measure the disruptiveness of papers. The first benchmark measure is the $DI^*$ (74, 79). The $DI^*$ is similar to the CD-index formula, but the numerator is $DC$ instead of $DC - CC$. This modification provides a more precise and consistent description of the level of disruptiveness.

$$DI^* = \frac{DC}{DC+CC+RC} \tag{1}$$

The second benchmark measure that we utilized is the simple DI (23). The simple DI is a variant of the CD-index that only considers the ratio of Disruptive Citation to the sum of disruptive and consolidating citations. This provides a simpler way to evaluate the degree of disruption caused by a citing paper.

$$simple\ DI = \frac{DC}{DC+CC} \tag{2}$$

The third benchmark measure combines both citation and disruption percentiles, denoted as C-CD P. This measure allows for a comparison between our proposed metric and existing approaches that consider both citation and disruption, as discussed in previous studies (78). The citation percentile (CP) and CD-index percentile (CDP) of papers are represented by CP and CDP, respectively. The formula for the C-CD P measure is as: $C - CD\ P = CP + CDP$. The illustrations and comparisons of some benchmarks are shown in Fig. A2.

To examine the consistency between DC and benchmark metrics, we utilized the Kendall's Tau coefficient (102), a nonparametric measure of association. The results are presented in Fig. A3. The analysis revealed that the CD-index, DI*, and simple DI metrics exhibited a negative correlation with citation (C) for Nobel-winning papers. This negative association persisted, albeit to a lesser extent, for the control group papers. These findings suggest that these benchmark metrics may not adequately capture the impact of papers, particularly when compared to citation. Furthermore, the inter-correlations among CD-index, DI*, simple DI, and C-CD P were consistently high across all three fields and for both Nobel-winning papers and control groups, with correlation values ranging from 0.64 to 0.96. This suggests that these metrics may be measuring similar aspects of a paper's disruptiveness, without providing distinct insights into the disruptive impact of papers. In contrast, the association between DC and C was remarkably high across all three fields and for both Nobel-winning papers and control groups, with correlation values ranging from 0.82 to 0.92. It is worth noting that this correlation was higher for Nobel-winning papers compared to their respective control groups, indicating that Nobel-winning papers tend to have a higher Disruptive Citation ratio compared to the control groups. However, the correlations between DC and the benchmark metrics for disruptive impact (CD-index, DI*, and simple DI) were relatively low. Interestingly, in the fields of Physics and Chemistry, these correlations were negative for Nobel-winning papers but positive for control groups. Similar patterns were observed for C-CD P. This discrepancy in correlation patterns suggests that the disruptive impact of scientific breakthroughs may be characterized by different underlying mechanisms compared to the control groups.

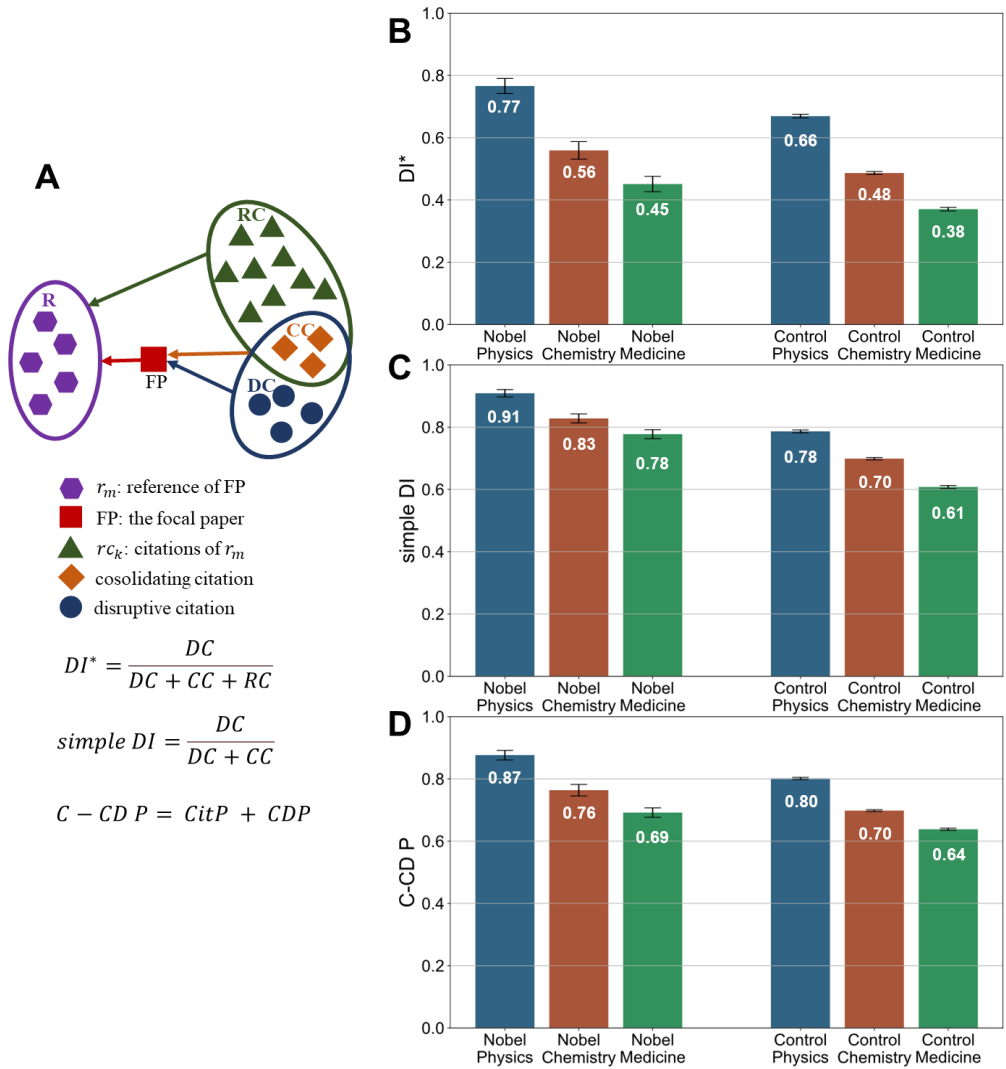

**Fig. A2. Illustration and comparison of DI\* and simple DI for Nobel-winning papers and control groups**. **(A)** Schematic representation of DI\* and simple DI of the focal paper, and their computational methodology. **(B)** Comparison of DI\* for Nobel-winning papers and control groups in each field. **(C)** Comparison of simple DI for Nobel-winning papers and control groups in each field. **(D)** Comparison of C-CD P for Nobel-winning papers and control groups in each field.

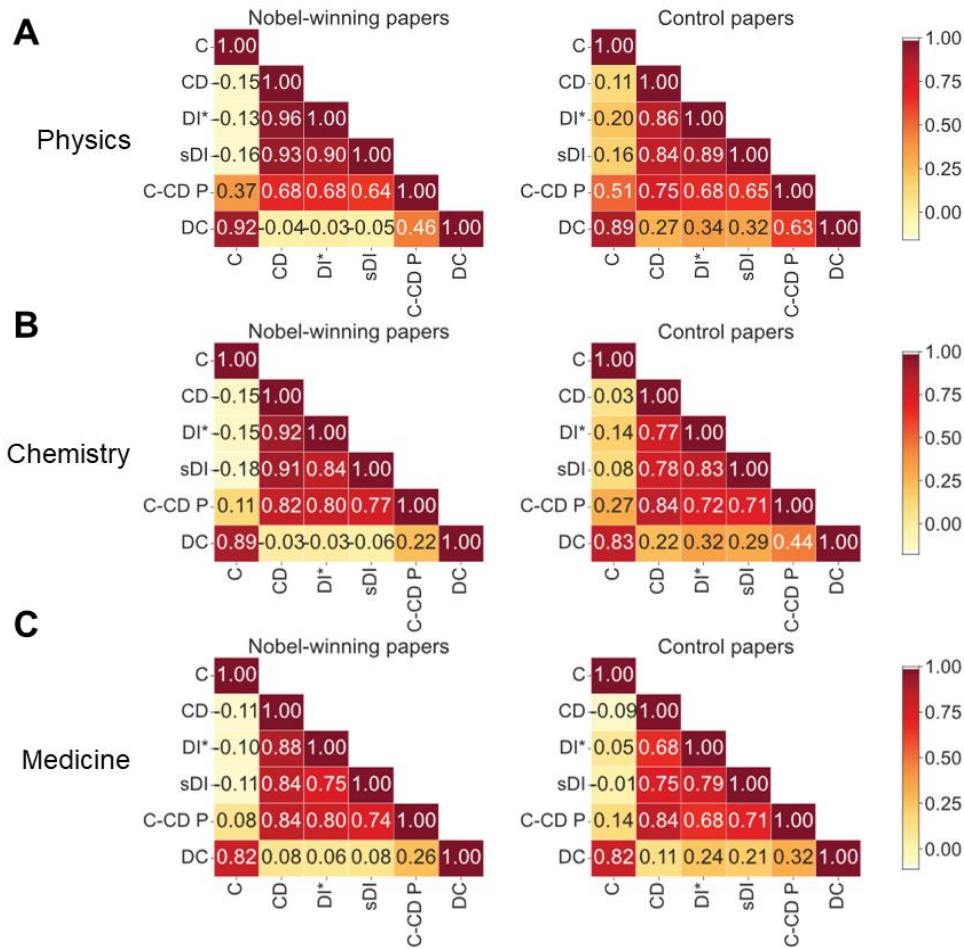

**Fig. A3. Correlation coefficients between citation (C), Disruptive Citation (DC) and other benchmark disruptive measures for Nobel-winning papers and control groups**, in **(A)** Physics, **(B)** Chemistry, and **(C)** Medicine.

**Note 4 Effectiveness and predictive powers**

In order to validate the effectiveness of DC, a comprehensive examination of effectiveness was conducted, comparing them with established benchmark metrics. First, to corroborate the potency of DC, we employed the Average Ranking (AR) of all Nobel-winning papers as a yardstick to gauge their convergent validity (33, 103). A diminished AR value signifies a more robust convergent validity, thus highlighting the metric's capacity to accurately represent the intended construct. As shown in Fig. A3, when using the average ranking (AR), the results highlight DC as a more effective approach in identifying groundbreaking research compared to other metrics. The average ranking of Nobel-winning papers using DC is impressive, with scores of 5.4% in Physics, 8.1% in Chemistry, and 6.4% in Medicine. The superior performance of DC across diverse disciplines underscores its effectiveness in capturing the essence of Nobel-winning contributions. In comparison, when employing citation (C), the average rankings of Nobel-winning papers are 5.8% in Physics, 9.2% in Chemistry, and 7.2% in Medicine. Although the citation's capacity to discern scientific breakthroughs across scientific domains is evident, DC demonstrates a marginally higher precision in measuring the impact of Nobel-winning papers. As predicting scientific breakthroughs is a crucial and tough issue in science, this clear superiority of DC has significant implications. In contrast, the CD-index exhibits diminished convergent validity, with average rankings of 54.25% in Physics, 52.42% in Chemistry, and 43.51% in Medicine. These findings suggest that the CD-index may not be as reliable a measure of research quality as DC. While the DI* and simple DI afford some valuable insights, they are also evidently less effective than the DC in encapsulating the true caliber of Nobel-winning papers.

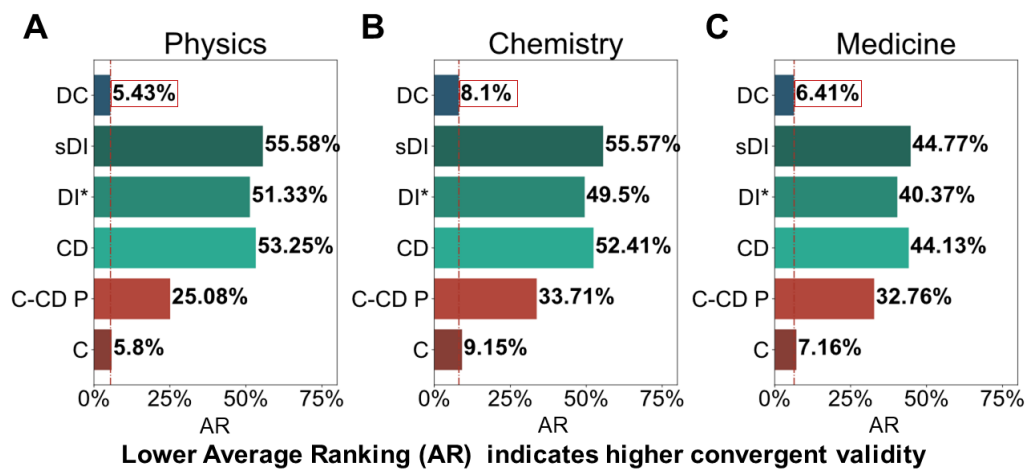

**Fig. A3. Convergent validity: the average ranking (AR) of Nobel-winning papers under each metric**, in **(A)** Physics, **(B)** Chemistry, and **(C)** Medicine.

Second, we meticulously assessed the convergent validity of various scientific metrics by employing the identification proportion (IP) as another robust criterion. The IP is defined as the ratio of Nobel-winning papers found within the top fraction of all sample data, as ranked by a specific metric under investigation (104). To compute the IP for each metric, we considered a wide range of ranked papers, ranging from the top 1% to the top 100%, and determined the average identification proportion for each metric. Fig. A4 presents a thorough analysis of the convergent validity, stratified by discipline, for each evaluated metric. Our findings demonstrate that Disruptive Citation (DC) exhibited the most exceptional performance across all disciplines, with an average identification proportion (IP) of 0.951 for Physics, 0.922 for Chemistry, and 0.941 for Medicine. These results highlight the potency of DC in identifying groundbreaking research and reaffirm its effectiveness as a metric for evaluating the impact of Nobel-winning papers. In contrast, the CD-index revealed markedly lower identification proportion (IP) values, with 0.457 in Physics, 0.475 in Chemistry, and 0.566 in Medicine. Similarly, DI* exhibited suboptimal performance, with average IP values of 0.477 for Physics, 0.516 for Chemistry, and 0.603 for Medicine. Lastly, simple DI exhibited the least favorable results, with the average IP values of 0.434 for Physics, 0.461 for Chemistry, and 0.56 for Medicine.

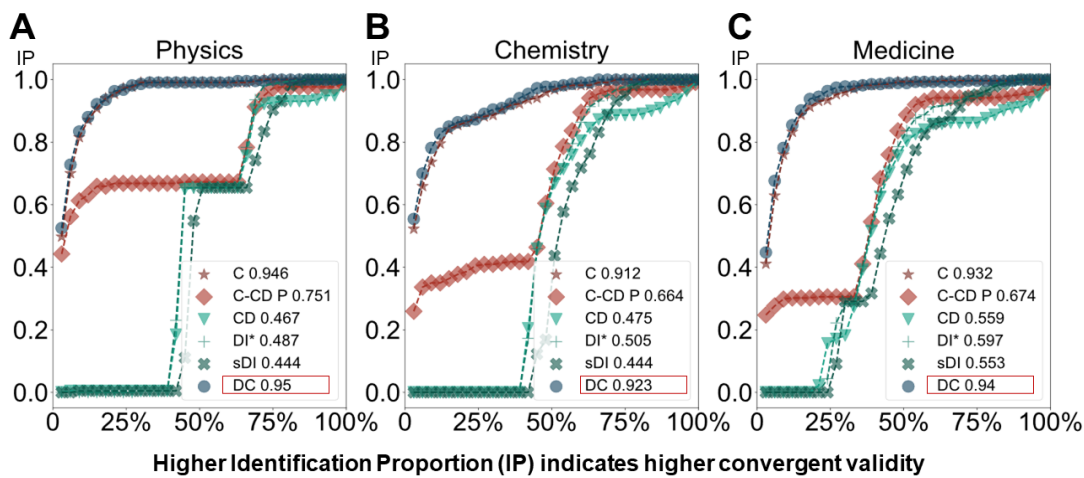

**Fig. A4. Convergent validity: the identification proportion (IP) of Nobel-winning papers under each metric**, in **(A)** Physics, **(B)** Chemistry, and **(C)** Medicine.

Third, we proceeded to investigate the classification task utilizing each metric and assessed the precision and recall values, subsequently calculating the F1 score. It is important to note that the classification task involved the utilization of a parameter, $k$, which represents the number of true values under each metric. In order to conduct a comprehensive comparison among the metrics, we set $k$ to range from 0 to the maximum value, which corresponds to the sample size, and calculate the average value as the final result. The classification task was performed within each field, and the outcomes are presented in Table A2. The results demonstrate the superior performance of DC in terms of precision, recall, and F1 score across all three fields. DC consistently outperforms the other benchmark measures, showcasing its effectiveness in accurately classifying papers and identifying groundbreaking research. This indicates that DC has a higher ability to correctly identify and capture the true positives, resulting in a lower number of false negatives.

These findings highlight the robustness and efficacy of DC compared to the benchmark measures. Our analysis underscores the importance of selecting appropriate metrics for evaluating the impact of scientific breakthroughs, particularly those that receive Nobel Prize accolades. The DC appears to offer a more reliable approach to capturing the transformative impact of scientific breakthroughs, which can help foster a deeper appreciation of the multifaceted nature of knowledge production and dissemination.

**Table A2. Classification task results.**

| Fields | Physics | | | Chemistry | | | Medicine | | |
|---|---|---|---|---|---|---|---|---|---|
| | Avg Precision | Avg Recall | Avg F1 | Avg Precision | Avg Recall | Avg F1 | Avg Precision | Avg Recall | Avg F1 |
| Citation (C) | 0.131 | 0.942 | 0.188 | 0.086 | 0.909 | 0.129 | 0.145 | 0.928 | 0.207 |
| CD-index | 0.024 | 0.467 | 0.046 | 0.016 | 0.476 | 0.031 | 0.039 | 0.559 | 0.072 |
| DI* | 0.025 | 0.487 | 0.047 | 0.017 | 0.505 | 0.033 | 0.041 | 0.596 | 0.076 |
| simple DI | 0.022 | 0.444 | 0.042 | 0.014 | 0.444 | 0.028 | 0.037 | 0.552 | 0.068 |
| C-CD P | 0.104 | 0.749 | 0.145 | 0.053 | 0.663 | 0.077 | 0.085 | 0.672 | 0.120 |
| DC | **0.134** | **0.946** | **0.191** | **0.089** | **0.919** | **0.132** | **0.150** | **0.936** | **0.212** |